\documentstyle[preprint,aps,graphicx,floats]{revtex}

\begin{document}

\preprint{$
\begin{array}{l}
\mbox{CERN-TH/2003-069}\\[-3mm]
\mbox{UB-HET-03-01}\\[-3mm]
\mbox{DESY-03-035}\\[-3.mm]
\mbox{April~2003} \\ [3mm]
\end{array}
$}

\title{Examining the Higgs boson potential at lepton and hadron
colliders: a comparative analysis}

\author{U.~Baur\footnote{e-mail: baur@ubhex.physics.buffalo.edu\\[-12.mm]}}
\address{Department of Physics,
State University of New York, Buffalo, NY 14260, USA\\[-2.mm]}
\author{T.~Plehn\footnote{e-mail: tilman.plehn@cern.ch\\[-12.mm]}}
\address{CERN Theory Group, CH-1211 Geneva 23, Switzerland\\[-2.mm] }
\author{D.~Rainwater\footnote{e-mail: david.rainwater@desy.de}}
\address{DESY Theorie, Notkestrasse 85, D-22603 Hamburg,
Germany\\[-3.mm]} 

\maketitle 

\begin{abstract}
\baselineskip13.pt  
We investigate inclusive Standard Model Higgs boson pair production at
lepton and hadron colliders for Higgs boson masses in the range
$120~{\rm GeV}\leq m_H\leq 200$~GeV. For $m_H\leq 140$~GeV we find
that hadron colliders have a very limited capability to determine the
Higgs boson self-coupling, $\lambda$, due to an overwhelming
background.  We also find that, in this mass range, supersymmetric
Higgs boson pairs may be observable at the LHC, but a measurement of
the self coupling will not be possible.  For $m_H > 140$~GeV we
examine $ZHH$ and $HH\nu\bar\nu$ production at a future $e^+e^-$ linear
collider with center of mass energy in the range of $\sqrt{s}=0.5 -
1$~TeV, and find that this is likely to be equally difficult.
Combining our results with those of previous literature, which has
demonstrated the capability of hadron and lepton machines to determine
$\lambda$ in either the high or the low mass regions, we establish a
very strong complementarity of these machines.
\end{abstract}

\newpage


\tightenlines

\section{Introduction}
\label{sec:sec1}

The CERN Large Hadron Collider (LHC) is widely regarded as capable of
directly observing the agent responsible for electroweak symmetry
breaking and fermion mass generation. This is generally believed to be
a light Higgs boson with mass $m_H<200$~GeV~\cite{lepewwg}. The LHC
will easily find a light Standard Model (SM) Higgs boson with very
moderate luminosity~\cite{wbf_ww,wbf_exp}. Moreover, the LHC will have
significant capability to determine many of its
properties~\cite{kunszt,tdr+,cms_tdr,lhc_nnlo}, such as its decay
modes and couplings~\cite{wbf_ll,Hcoup,Yb,Yt,tth_nlo}, including
invisible decays\cite{wbf_inv} and possibly even rare decays to light
fermions~\cite{Hmumu}. An $e^+e^-$ linear collider with a center of
mass energy of 350~GeV or more can significantly improve these
preliminary measurements, in some cases by an order of magnitude in
precision, if an integrated luminosity of 500~fb$^{-1}$ can be
achieved~\cite{LC}.

Starting from the requirement that the Higgs boson has to restore
unitarity of weak boson scattering at high energies in the
SM~\cite{unit}, perhaps the most important measurement after the Higgs
boson discovery is of the Higgs potential itself, which requires
measurement of the trilinear and quartic Higgs boson self-couplings,
$\lambda$ and $\tilde\lambda$, respectively.  These can be probed
directly only by multiple Higgs boson production (at any future
collider). While $\lambda$ can be measured in Higgs boson pair
production, triple Higgs boson production is needed to probe
$\tilde\lambda$. Since the cross sections for three Higgs boson
production processes are more than a factor $10^3$ smaller than those
for Higgs pair production at linear colliders~\cite{LC_HH3,H3}, and
about an order of magnitude smaller at hadron
colliders~\cite{higgs_self,lhc_hh}, the quartic Higgs boson coupling
will likely remain elusive even at the highest collider energies and
luminosities considered so far.\medskip

Several studies of Higgs boson pair production in $e^+e^-$ collisions
have been conducted over the past few
years~\cite{LC_HH3,LC_HH1,LC_HH2,LC_HH4}, deriving quantitative
sensitivity limits for the trilinear Higgs self-coupling $\lambda$ for
several proposed linear colliders with center of mass energies
spanning the range from 500~GeV to 3~TeV. For example, a study
employing neural net techniques found that $\lambda$ could be measured
for $m_H=120$~GeV with a precision of about $20\%$ at a 500~GeV linear
collider with an integrated luminosity of 1~ab$^{-1}$~\cite{LC_HH4}.
In contrast, the potential of the LHC, a luminosity-upgraded LHC
(SLHC) which would gather 10~times the amount of data expected in the
first run, and a Very Large Hadron Collider (VLHC), has been examined
only recently~\cite{SLHC,BPR,blondel}. These studies investigated
Higgs pair production via gluon fusion and subsequent decay to
same-sign dileptons and three leptons via $W$ bosons. They established
that future hadron machines can probe the Higgs potential for
$m_H>150$~GeV. At the LHC, with an integrated luminosity of
300~fb$^{-1}$, a vanishing of $\lambda$ can be excluded at the $95\%$
confidence level or better over the entire range $150~{\rm
  GeV}<m_H<200$~GeV. At a VLHC, the Higgs boson self-coupling can be
determined with a precision of a few percent with the same integrated
luminosity for $m_H=180$~GeV, which is similar or better than the
limits achievable at a 3~TeV $e^+e^-$ collider with
5~ab$^{-1}$~\cite{LC_HH3}.\medskip

In this paper we present an analysis of the converse: we look at Higgs
boson pair production for $m_H\leq 140$~GeV at future hadron
colliders, and estimate the prospects for probing $\lambda$ if
$m_H\geq 150$~GeV at a future linear collider with a center of mass
energy of $0.5-1$~TeV. To fully compare the capabilities of $e^+e^-$
linear colliders and hadron colliders, we also extrapolate the results
of Ref.~\cite{LC_HH4} to $m_H>120$~GeV and center of mass energies
larger than 500~GeV. In Sec.~\ref{sec:sec2} we recall the definition
of the Higgs boson self-couplings and briefly discuss SM and non-SM
predictions for these parameters. In Sec.~\ref{sec:lhc} we analyze
Higgs boson pair production via gluon fusion with subsequent decay
into four $b$ jets and $b\bar{b}\tau\tau$ final states at the LHC,
SLHC and a VLHC, which we assume to be a $pp$ collider operating at
200~TeV with a luminosity of ${\cal L}=2\times 10^{34}~{\rm cm^{-2}\,
  s^{-1}}$~\cite{vlhc}. We also briefly comment on the prospects for
observing a pair of MSSM Higgs bosons in the $b\bar{b}\tau\tau$ decay
channel.  We discuss Higgs boson pair production in $e^+e^-$
collisions in Sec.~\ref{sec:lc}. In Sec.~\ref{sec:pot} we determine
how well the Higgs potential could be reconstructed at future lepton
and hadron colliders. We draw conclusions in Sec.~\ref{sec:conc}.

\section{Higgs boson self-couplings}
\label{sec:sec2}

The trilinear and quartic Higgs boson couplings $\lambda$ and
$\tilde\lambda$ are defined through the potential
\begin{equation}
\label{eq:Hpot}
V(\eta_H) \, = \, 
{1\over 2}\,m_H^2\,\eta_H^2\,+\,\lambda\, v\,\eta_H^3\,+\,{1\over 4}\,
\tilde\lambda\,\eta_H^4 ,
\end{equation}
where $\eta_H$ is the physical Higgs field, $v=(\sqrt{2}G_F)^{-1/2}$
is the vacuum expectation value, and $G_F$ is the Fermi constant. In
the SM,
\begin{equation}
\label{eq:lamsm}
\tilde\lambda=\lambda=\lambda_{SM}={m_H^2\over 2v^2}\,.
\end{equation}

Regarding the SM as an effective theory, the Higgs boson
self-couplings $\lambda$ and $\tilde\lambda$ are {\it per se} free
parameters.  $S$-matrix unitarity constrains $\tilde\lambda$ to
$\tilde\lambda\leq 8\pi/3$~\cite{unit}.  Since future collider
experiments likely cannot probe $\tilde\lambda$, we concentrate on the
trilinear coupling $\lambda$ in the following. The quartic Higgs
coupling does not affect the Higgs pair production processes discussed
in this paper. Our results, with the exception of the constraints on
$V(\eta_H)$ discussed in Sec.~\ref{sec:pot} (where we assume
$\tilde\lambda=\lambda_{SM}$) are therefore independent of the value
assumed for~$\tilde\lambda$.\medskip

In the SM, radiative corrections decrease $\lambda$ by $4-11\%$ for
$120~{\rm GeV}< m_H<200$~GeV~\cite{yuan}. Larger deviations are
possible in scenarios beyond the SM. For example, in two Higgs doublet
models where the lightest Higgs boson is forced to have SM like
couplings to vector bosons, quantum corrections may increase the
trilinear Higgs boson coupling by up to $100\%$~\cite{yuan}. In the
Minimal Supersymmetric Standard Model (MSSM), loop corrections modify
the self-coupling of the lightest Higgs boson, which has SM-like
couplings, by up to $8\%$ for light stop squarks~\cite{hollik}.
Anomalous Higgs boson self-couplings also appear in various other
scenarios beyond the SM, such as models with a composite Higgs
boson~\cite{georgi}, or in ``little Higgs'' models~\cite{lhiggs}. In
many cases, the anomalous Higgs boson self-couplings can be
parameterized in terms of higher dimensional operators which are
induced by integrating out heavy degrees of freedom. A systematic
analysis of Higgs boson self-couplings in a higher dimensional
operator approach can be found in Ref.~\cite{tao}.

\section{A low mass Higgs boson at hadron colliders}
\label{sec:lhc}

At LHC energies, inclusive Higgs boson pair production is dominated by
gluon fusion. Other processes, such as weak boson fusion, $qq\to
qqHH$~\cite{wbf}, associated production with heavy gauge bosons,
$q\bar{q}\to VHH$ ($V=W,\,Z$)~\cite{assoc}, or associated production
with top quark pairs, $gg,\,q\bar{q}\to t\bar{t}HH$~\cite{SLHC}, yield
cross sections which are factors of 10 --~30 smaller than that for
$gg\to HH$~\cite{SLHC,lhc_hh}. Since Higgs boson pair production at
the LHC is already rate limited, we concentrate on the gluon fusion
process in the following. For $m_H<140$~GeV, the dominant decay mode
of the SM Higgs boson is $H\to b\bar{b}$. In the following, we examine
the largest overall branching ratio production, which yields four
$b$-quark final states, and decays where one Higgs boson decays into
$b\bar{b}$ and the other into a $\tau$ pair, $gg\to HH\to
b\bar{b}\tau^+\tau^-$.\medskip

For all our calculations we assume an integrated luminosity of
300~fb$^{-1}$ for LHC and VLHC~\cite{vlhc}, and
3000~fb$^{-1}$\cite{SLHC} for the SLHC. We choose
$\alpha_s(M_Z)=0.1185$~\cite{long}, and assume a $b$-tagging
efficiency of $50\%$ for all hadron colliders. Signal and background
cross sections are consistently calculated using CTEQ4L~\cite{cteq}
parton distribution functions. We include minimal detector effects by
Gaussian smearing of the parton momenta according to ATLAS
expectations~\cite{tdr+}, and take into account energy loss in the
$b$-jets via a parameterized function. In addition, we include an
efficiency of $68\%$ for capturing each $H\to b\bar{b}$ or
$H\to\tau^+\tau^-$ decay in the signal in its relevant mass bin. All
tree level processes are calculated using {\sc
  madgraph}~\cite{madgraph} and retain a finite $b$-quark mass of
4.6~GeV.

\subsection{$\mathbf pp\to 4b$}

We perform the calculation of the signal, $gg\to HH\to 4b$, as in
Ref.~\cite{BPR}, including the effects of next-to-leading order (NLO)
QCD corrections via a multiplicative factor $K=1.65$ at LHC and
$K=1.35$ at VLHC energies~\cite{hh_nlo}.  The largest background to
consider is QCD continuum four $b$-quark production. The factorization
and renormalization scale choices are taken to be $m_H$. There is
large uncertainty due to scale variation in the QCD backgrounds, but
this is irrelevant given our findings which follow. The choice of
scales for the signal rate can have a large impact as well, for
example varying the scale between the Higgs boson mass and the
invariant mass of the final state Higgs boson pair, as chosen as a
default in the public version of the NLO matrix
elements~\cite{hh_nlo}.  However, after including the approximate NLO
corrections as a $K$-factor we find that the scale dependence of the
cross section as a measure of the theoretical uncertainty is strongly
reduced, which is the main reason to compute and include these higher
order corrections. We require that all four $b$-quarks in the event
are tagged.

The kinematic acceptance cuts for events at the LHC are:
\begin{eqnarray}
\label{eq:cuts1}
\nonumber &
p_T(b) > 75,\, 65,\, 40,\, 20~{\rm GeV} \; , \qquad
|\eta(b)| < 2.5 \; , \\
&
\Delta R(b,b) > 0.7 \; , \qquad 
m_H-30~{\rm GeV} \, < \, m_{b\bar{b}} \, < \, m_H+10~{\rm GeV} 
\end{eqnarray}
which are motivated first by requirements that these all-hadronic
events can pass the ATLAS and CMS triggers with reasonable
efficiency~\cite{jakobs}, and that two $b$-quark pairs each
reconstruct to a window around the known Higgs boson mass, asymmetric
due to energy loss in the $b$-jets. This invariant mass constraint on
the six possible bottom pairs defines the candidates to reconstruct
the Higgs bosons.  We also use the cuts of Eq.~(\ref{eq:cuts1}) for
the SLHC and VLHC. Preliminary studies concluded~\cite{SLHC,snowm}
that cuts similar to those listed in Eq.~(\ref{eq:cuts1}) should be
sufficient, although increased background from event pile-up is
expected to degrade detector performance at the SLHC.

Comparing the signal and the backgrounds, there is first of all an
important difference between the $4b$ final state and the $4W$ final
state investigated previously~\cite{SLHC,BPR,blondel}: the background
of the bottom final state does not involve any massive particles, like
top quarks or $W$ bosons. All four bottom jets in the QCD background
process are either produced without a strong azimuthal correlation
between each other or come from (mostly collinear) gluon splitting.
The latter will to a large degree be removed by the $m_{b\bar{b}}$ cut
together with the $\Delta R(b,b)$ cut. This on the one hand requires a
hard central gluon to split into two bottom quarks, which on the other
hand cannot be boosted together. Even though the Higgs bosons are
produced close to rest, they are still very massive states with a
non-negligible transverse momentum, which decay to effectively
massless bottom quarks. Translated into the geometry of the four
bottom jets, this means that we can require a sizable transverse
momentum of the bottom pairs which should reconstruct the Higgs
bosons, and also require that these bottom jets lie close to each
other in the azimuthal plane:
\begin{eqnarray}
\label{eq:cuts2}
\nonumber &
p_T(bb)^{\rm min} > 105~{\rm GeV} \; , \qquad
p_T(bb)^{\rm max} > 115~{\rm GeV} \; , \\
&
\triangle\phi(b,b)^{\rm min} < 0.5\pi \; , \qquad
\triangle\phi(b,b)^{\rm max} < 0.7\pi.
\end{eqnarray}
As in the $4W$ signal case~\cite{BPR}, we will later try to determine
the Higgs boson self-coupling from the shape of the invariant mass of
the final state. For that reason we do not apply any cuts which make
use of the fact that the signal involves two heavy massive particles
produced in a fairly narrow range of the $4b$ invariant mass. However,
for a fixed invariant mass of the $4b$ final state we expect more
forward jets for the QCD background which in turn do not need to have
as large a transverse momentum. We therefore require the scalar sum of
the transverse momentum to obey:
\begin{equation}
\label{eq:cuts3}
\sum{p}_T > 270~{\rm GeV}.
\end{equation} \medskip

\begin{figure}[t!] 
\begin{center}
\includegraphics[width=13.3cm]{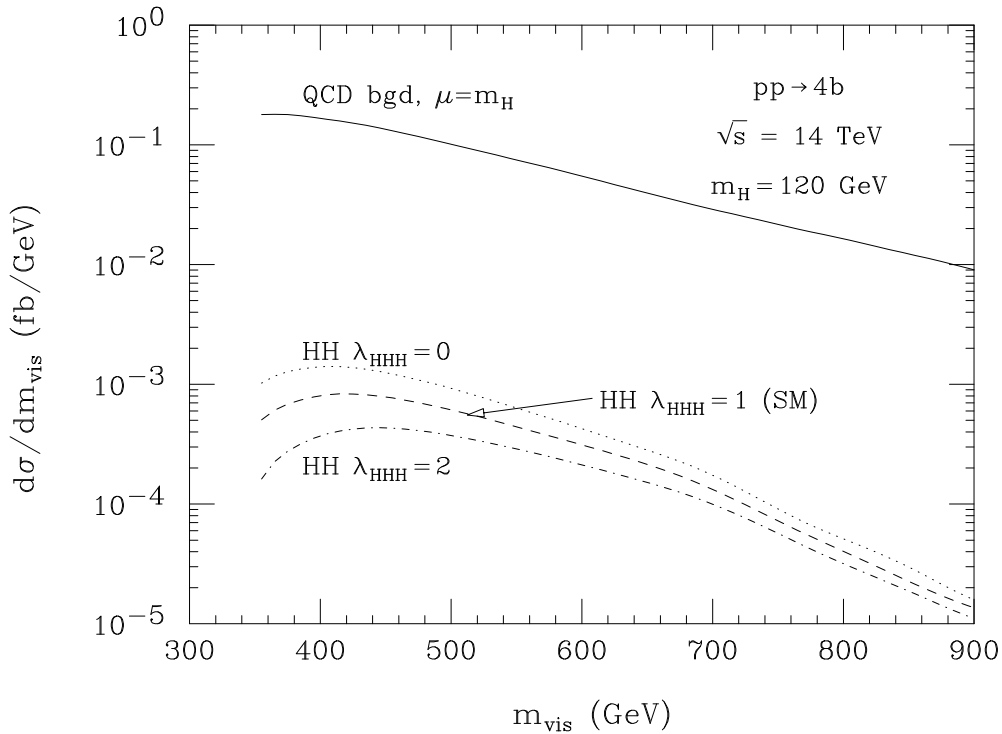}
\vspace*{2mm}
\caption[]{\label{fig:m4b} 
  Distribution of the visible invariant mass, $m_{vis}$, in $pp\to
  4b$, after all kinematic cuts (Eqs.~(\protect\ref{eq:cuts1})
  --~(\protect\ref{eq:cuts3})), for the QCD continuum background
  (solid) and the SM signal for $m_H=120$~GeV (dashed) at the LHC. The
  dotted and dot-dashed lines show the signal cross section for
  $\lambda_{HHH}=\lambda/\lambda_{SM}=0$ and $\lambda_{HHH}=2$,
  respectively.} 
\vspace{-7mm}
\end{center}
\end{figure}

Extracting the Higgs boson self-coupling follows the same path as for
the $4W$ final state~\cite{BPR}. To discriminate signal and
background, we can use the visible invariant mass, $m_{vis}$, which
for the $4b$ final state is the invariant mass of the Higgs boson
pair, corrected for energy loss of the $b$ jets. The $m_{vis}$
distributions of the signal for $m_H=120$~GeV and the QCD background
at the LHC are shown in Fig.~\ref{fig:m4b}.
Even after all cuts, the QCD background is more than two orders of
magnitude larger than the signal. For an integrated luminosity of
300~fb$^{-1}$, about 55~signal events are expected in the SM. The
absence of a Higgs boson self-coupling
($\lambda_{HHH}=\lambda/\lambda_{SM}=0$) results in a Higgs boson pair
production cross section about a factor~1.6 larger than the SM result,
whereas increasing $\lambda$ to twice the SM value decreases the rate
by a factor~1.7. At VLHC energies, the cross sections of the signal
and background are a factor 100 and 60 larger than those at the LHC,
giving negligible improvement in the signal to background ratio,
$S/B$.

The small signal cross section combined with the huge QCD $4b$
background make it essentially impossible to determine the Higgs boson
self-coupling in $pp\to 4b$.  We quantify this statement by performing
a $\chi^2$ test on the $m_{vis}$ distribution, similar to that
described in Ref.~\cite{BPR}.  Except for the Higgs boson
self-coupling, we assume the SM to be valid. To approximately take
into account the unknown NLO QCD corrections to $pp\to 4b$, we
multiply the QCD $4b$ differential cross section by a uniform
$K$-factor of $K=1.3$ and allow for a normalization uncertainty of
$10\%$ for the SM cross section. For $m_H=120$~GeV we then obtain
$1\sigma$ bounds of
\begin{eqnarray}
-6.8<&\Delta\lambda_{HHH}<10.1 \qquad &({\rm LHC}), \nonumber \\ 
-3.1<&\Delta\lambda_{HHH}<6.0 \qquad  &({\rm SLHC}), \nonumber \\
-1.3<&\Delta\lambda_{HHH}<2.4 \qquad  &({\rm VLHC}),
\label{eq:limit1}
\end{eqnarray}
where
\begin{equation}
\Delta\lambda_{HHH}={\lambda\over\lambda_{SM}}-1.
\end{equation}\medskip

For comparison, a 500~GeV linear collider with an integrated
luminosity of 1~ab$^{-1}$ could determine $\lambda$ with a precision
of about $20\%$ for $m_H=120$~GeV~\cite{LC_HH4}.  For $m_H>120$~GeV,
the $H\to b\bar{b}$ branching ratio drops quickly.  Since the
background cross section decreases only slightly, $S/B$, and thus the
bounds on the $\lambda$, worsen with increasing values of $m_H$.

\subsection{$\mathbf pp\to b\bar{b}\tau^+\tau^-$}

The insensitivity of $4b$ production to the Higgs boson self-coupling
is largely due to the overwhelming QCD background. A more advantageous
$S/B$ is conceivable if one of the Higgs bosons in $gg\to HH$ decays
into a $\tau$ pair. In this case, the main contributions to the
background arise from continuum $b\bar{b}\tau^+\tau^-$ and
$t\bar{t}\to W^+W^-b\bar{b}\to\tau^+\nu_\tau\tau^-\bar\nu_\tau
b\bar{b}$ production. We calculate both processes using tree level
matrix elements which include all decay correlations. Top quarks are
generated on-shell. The calculation of the signal proceeds as for the
$4b$ final state. We assume that both $b$-quarks are tagged.  Because
of its small mass, we simulate $\tau$ decays in the collinear
approximation. All $\tau$ decays are calculated following the approach
described in Ref.~\cite{wbf_ll}.

If both $\tau$ leptons decay leptonically, $4b$ production where two
$b$-quarks decay leptonically represents an additional potentially
large background, even if one requires that both leptons are isolated.
In addition, the signal cross section is suppressed by the small
branching ratio of about $13\%$ if both $\tau$ leptons decay
leptonically. In the following we therefore only consider decays where
at least one $\tau$ lepton decays hadronically.

To identify $b\bar{b}\tau\tau$ events in a hadron collider
environment, one has to trigger on the $\tau$ pair. At a luminosity of
${\cal L}=10^{34}~{\rm cm^{-2}\,s^{-1}}$ this requires severe
transverse momentum cuts on the $\tau$ decay
jet~\cite{wbf_exp,sasha,cavalli}.  To ensure that an event is
successfully recorded in which one $\tau$ lepton decays leptonically
($\tau\to\ell\nu_\ell\nu_\tau$, $\ell=e,\,\mu$) and the other
hadronically ($\tau\to h\nu_\tau$), we impose the following transverse
momentum and rapidity cuts on the $\tau$ decay products~\cite{sasha}:
\begin{eqnarray}
& p_T(\ell)>20~{\rm GeV}, \qquad |\eta(\ell)|<2.5, \nonumber \\ 
& p_T(h)>50~{\rm GeV}, \qquad |\eta(h)|<2.5, \nonumber \\ 
& \Delta R(\ell,h)>0.4, \qquad \Delta R(\ell,b)>0.4, 
\qquad \Delta R(h,b)>0.4.
\label{eq:cuts10}
\end{eqnarray}
We assume that hadronically decaying $\tau$-jets which satisfy
Eq.~(\ref{eq:cuts10}) will be identified with an efficiency of
$\epsilon_\tau=0.33$ and discriminated from other jets with a
rejection factor of 500 or more~\cite{jakobs}. The large rejection
factor makes the $b\bar{b}jj$ background, where the two non-$b$ jets
fake hadronically decaying $\tau$ leptons, negligible.  If both $\tau$
leptons decay hadronically, even more severe $p_T$ cuts are
required~\cite{sasha}:
\begin{eqnarray}
& p_T(h_{1,2})>65~{\rm GeV}, \qquad |\eta(h_{1,2})|<2.5 \nonumber \\
& \Delta R(h_1,h_2)>0.6, \qquad \Delta
R(h_{1,2},b)>0.4,
\label{eq:tau1}
\end{eqnarray}
where $h_{1,2}$ are the $\tau$ decay jets. 

For the signal, the $\tau$-pair invariant mass can be reconstructed
from the observable $\tau$ decay products and the missing transverse
momentum vector of the event~\cite{soldate}. To reduce the background,
we therefore impose a cut on the reconstructed $\tau$-pair invariant
mass,
\begin{equation}
m_H-2\Delta \, < \, m_{\tau\tau}^{\rm rec} \, < \, m_H+2\Delta ,
\end{equation}
where $\Delta$ is the $1\sigma$ half-width for the $H$ peak. $\Delta$
ranges from about 7.5~GeV for $m_H=120$~GeV to 15~GeV for
$m_H=140$~GeV. Finally, we impose the following cuts on the $b$-jets:
\begin{eqnarray}
& p_T(b)>30~{\rm GeV}, \qquad
|\eta(b)| <2.5, \nonumber \\ 
& \Delta R(b,b)>0.4, \qquad m_H-20~{\rm GeV} \, < \, m_{b\bar{b}} \,
< \, m_H+20~{\rm GeV} .
\label{eq:cuts11}
\end{eqnarray}
Unlike for the $4b$ final state, we have chosen a symmetric window
around $m_H$ for the $b\bar{b}$ invariant mass. We found that although
the energy loss in the $b$-quarks creates a non-Gaussian tail for
$m_{b\bar{b}}<m_H$, it makes little difference whether a symmetric or
asymmetric $m_{b\bar{b}}$ cut is imposed for $b\bar{b}\tau\tau$
production. Since there are more $b\bar{b}$ combinations possible, the
difference between a symmetric and asymmetric window for
$m_{b\bar{b}}$ is more pronounced for $4b$ production.  We also use
the cuts of Eqs.~(\ref{eq:cuts10}) and (\ref{eq:cuts11}) for the SLHC and
VLHC. We note that since the $p_T$ distributions of the $\tau$ decay
products fall steeply with increasing transverse momenta, the $gg\to
HH\to b\bar{b}\tau\tau$ cross section depends sensitively on the cuts
in Eqs.~(\ref{eq:cuts10}) and~(\ref{eq:tau1}).\medskip

As before, the effects of next-to-leading order (NLO) QCD corrections
are included in our calculation via multiplicative factors which are
summarized in Table~\ref{tab:one}. Note that this is one of the rare
instances where the NLO corrections are known for the signal and all
major backgrounds.  The factorization and renormalization scale
choices are taken to be $m_H$ for the signal and the
$b\bar{b}\tau\tau$ background; for $t\bar{t}$ production we choose the
top quark mass, $m_t$.\medskip

\begin{figure}[t!] 
\begin{center}
\includegraphics[width=13.3cm]{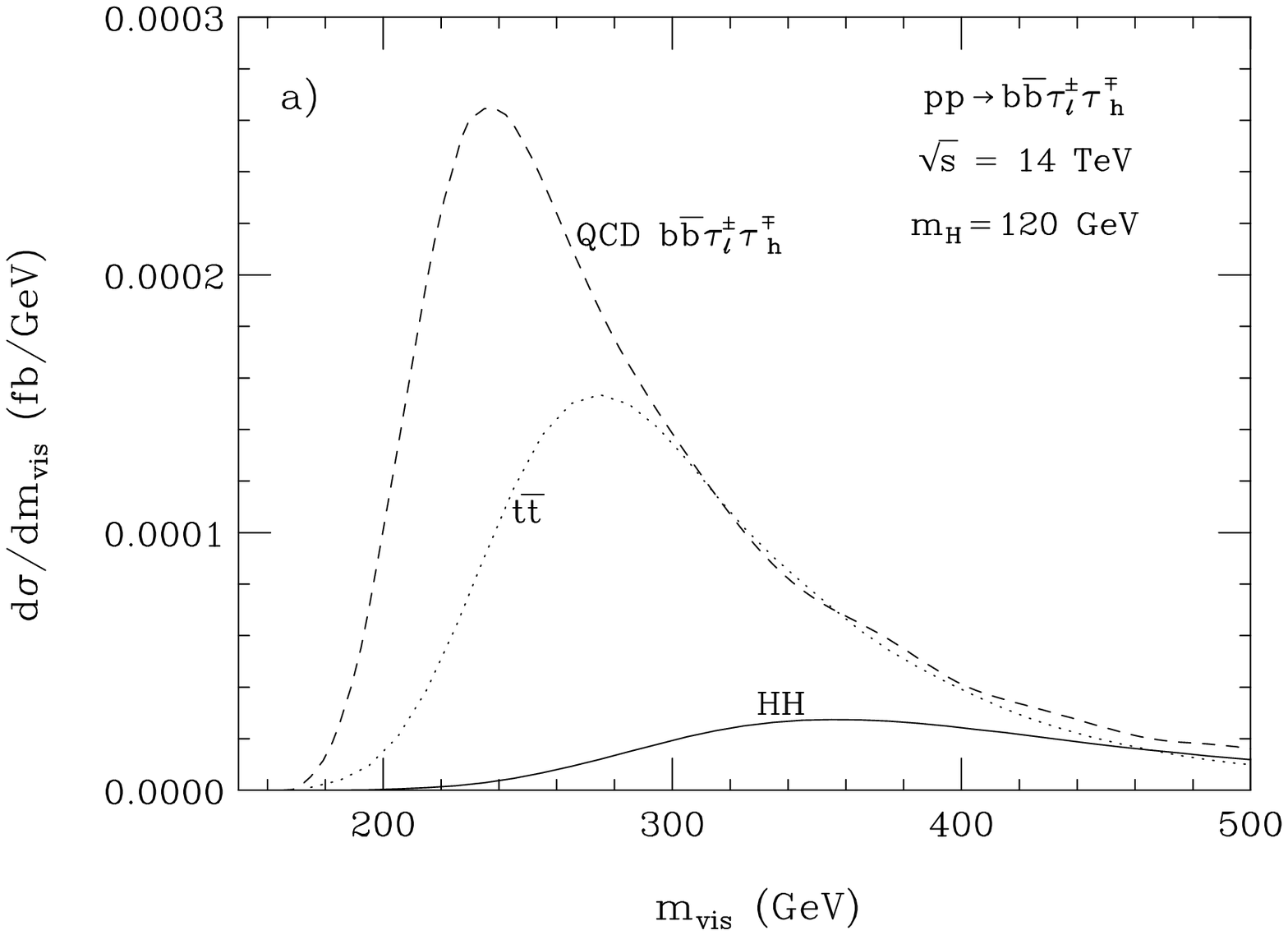}\\[3.mm]
\includegraphics[width=13.3cm]{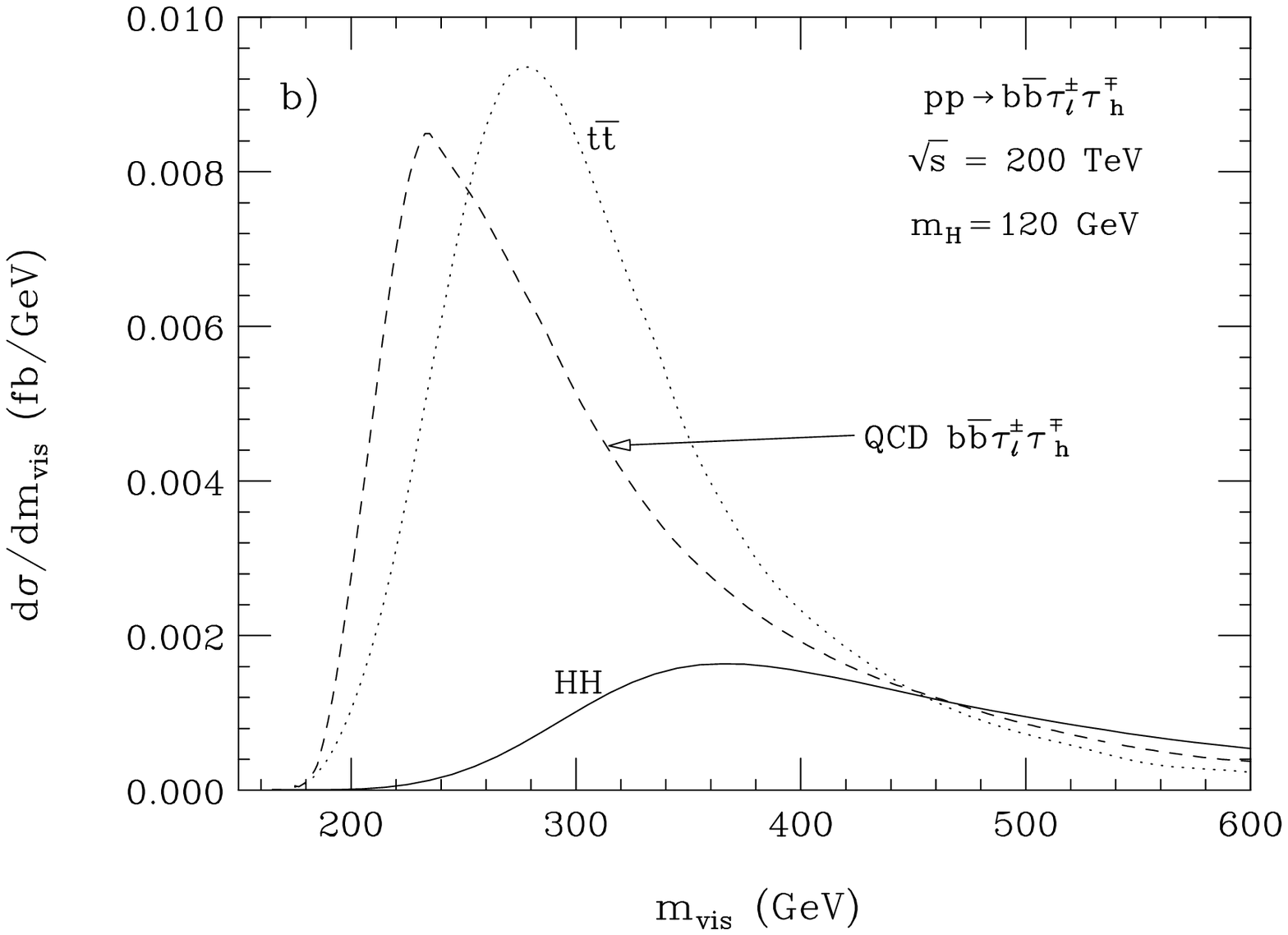}
\vspace*{2mm}
\caption[]{\label{fig:mbbtt} 
  Distribution of the visible invariant mass, $m_{vis}$, after all
  kinematic cuts, in $pp\to b\bar{b}\tau_\ell\tau_h$ for the SM signal
  with $m_H=120$~GeV (solid), the QCD continuum background (dashed)
  and the $t\bar{t}$ background (a) at the LHC, and (b) at the VLHC.
  $\tau_\ell$ ($\tau_h$) indicates that the $\tau$ lepton decays
  leptonically (hadronically).} \vspace{-7mm}
\end{center}
\end{figure}

\begin{table}
\caption{$K$-factors for $gg\to HH$~\protect\cite{hh_nlo}, 
$b\bar{b}\tau\tau$ production~\protect\cite{ellis}, and $t\bar{t}$ 
production~\protect\cite{lhctop}. The Higgs boson mass is assumed to 
be $m_H=120$~GeV. The factorization and renormalization scales used 
are described in the text.}   
\label{tab:one}
\vskip 3.mm
\begin{tabular}{cccc}
energy & $HH$ & $b\bar{b}\tau^+\tau^-$ & $t\bar{t}$ \\
\tableline
$\sqrt{s}=14$~TeV (LHC) & 1.65 & 1.21 & 1.35 \\
$\sqrt{s}=200$~TeV (VLHC) & 1.35 & 0.79 & 1.00
\end{tabular}
\end{table}
For the cuts of Eq.~(\ref{eq:cuts10}), the cross section for $pp\to
HH\to b\bar{b}\tau\tau$ where both $\tau$ leptons decay hadronically
is about a factor~7 smaller than that where one of them decays into
leptons. In the following we therefore consider the latter only. The
$HH$ signal in $pp\to b\bar{b}\tau_\ell\tau_h$ ($\tau_\ell$ and
$\tau_h$ denote the leptonically and hadronically decaying $\tau$
leptons, respectively) and the continuum $b\bar{b}\tau\tau$ and
$t\bar{t}$ backgrounds can again be discriminated using the visible
invariant mass distribution, $m_{vis}$. For the
$b\bar{b}\tau_\ell\tau_h$ final state, $m_{vis}$ is given by
\begin{equation}
m^2_{vis}=
\left [ E_b+E_{\bar{b}}+E_\ell+E_h \right ]^2-
\left [ 
\mathbf p_b + \mathbf p_{\bar{b}}+\mathbf p_\ell+\mathbf p_{had}
\right ]^2~,
\end{equation}
where $E$ and $\mathbf p$ denote the measured energy and momentum of a
particle.  Fig.~\ref{fig:mbbtt} demonstrates that, for $m_H=120$~GeV,
the signal peaks at significantly larger values of $m_{vis}$ than the
background processes.
The QCD $b\bar{b}\tau\tau$ background peaks at smaller $m_{vis}$
values because the $b\bar{b}$ system does not form a heavy resonance.
The $t\bar{t}$ background peaks below $2m_t$ due to the additional
neutrinos produced in the $W\to\tau\nu_\tau$ decays. Although the
shape of the visible invariant mass distribution provides a tool to
discriminate signal and background in $pp\to b\bar{b}\tau_\ell\tau_h$,
the combined QCD $b\bar{b}\tau\tau$ and $t\bar{t}$ background is much
larger than the signal. In addition, the signal cross section is very
small: at the SLHC (VLHC), one expects about 20 (140)~signal events
for $m_H=120$~GeV. The number of signal events decreases quickly with
increasing Higgs boson mass, due to the rapidly falling $H\to
b\bar{b}$ and $H\to\tau\tau$ branching fractions.  Since the
background decreases much less, $S/B$, and thus the bounds on
$\lambda$, worsen with increasing values of $m_H$.\medskip

To determine whether useful information on the Higgs boson
self-coupling can be extracted from the $b\bar{b}\tau_\ell\tau_h$
final state, we again perform a $\chi^2$ test on the $m_{vis}$
distribution. Since the signal cross section is too small to be
observable at the LHC, we derive bounds only for the SLHC and a VLHC.
Allowing for a normalization uncertainty of $10\%$ of the SM cross
section, for $m_H=120$~GeV we find $1\sigma$ bounds of
\begin{eqnarray}
-1.6 <&\Delta\lambda_{HHH}<3.1 \qquad  &({\rm SLHC}), \nonumber \\
-0.84<&\Delta\lambda_{HHH}<0.96 \qquad &({\rm VLHC}).
\end{eqnarray}
While these bounds are a factor 1.5~--~2.5 more stringent than those
which can be obtained from $HH\to 4b$ (see Eq.~(\ref{eq:limit1})),
they are a factor 5~--~10 less stringent than those one hopes to
achieve with 1~ab$^{-1}$ at a linear collider operating at
500~GeV~\cite{LC_HH4}. For $m_H=140$~GeV, we obtain limits which are
more than a factor two weaker than those for $m_H=120$~GeV.

\subsection{Supersymmetric Higgs Bosons}

We close this section with a brief remark on Higgs boson pair
production in supersymmetric models. In the MSSM, pair production of
$CP$-odd Higgs bosons, $gg\to AA$, is enhanced by a factor
$\tan^4\beta$~\cite{lhc_hh}. $AA$ production may thus be observable in
the $b\bar{b}\tau_\ell\tau_h$ final state at the LHC if $\tan\beta$ is
sufficiently large and $m_A$ is small.  In this region of
supersymmetric parameter space the $CP$-odd Higgs boson, $A$, can of
course be produced via bottom quark fusion, which is enhanced by a
factor $\tan^2 \beta$ compared to the usual gluon fusion process.
However, it is difficult to observe the additional final state
$b$-jets and to verify that the process indeed proceeds through an
enhanced bottom Yukawa coupling.  Observing the correspondingly huge
increase of the pair production cross section would confirm the
presence of a large $\tan \beta$ enhancement factor. For example, for
$m_A=120$~GeV and $\tan\beta=35$, we find a cross section (including
$b$-tagging and hadronic $\tau$ decay efficiencies, and using the same
cuts as in the SM case) of about 0.06~fb at the LHC, yielding about
20~signal events for an integrated luminosity of 300~fb$^{-1}$. The
combined $t\bar{t}$ and QCD $b\bar{b}\tau\tau$ background is about
17~events. $AA$ production thus should be observable at the LHC with a
significance of $5\sigma$ or more if $\tan\beta>35$. Unfortunately,
since the Feynman diagrams involving the $HAA$ and $hAA$ couplings are
only enhanced by a factor $\tan^2\beta$, the process $gg\to AA$ is
very insensitive to these couplings.

It should be noted that, for the values of $m_A$ and $\tan\beta$
chosen here, $A$ and the heavy $CP$ even Higgs boson, $H$, are almost
degenerate in mass ($m_H\approx 125$~GeV). The invariant mass
resolution of the LHC detectors will make it impossible to separate
the $A$ and $H$ bosons in this case, and one will in fact observe a
combined $AA+AH+HH$ signal. For $m_A=120$~GeV and $\tan\beta=35$, the
$HH$ cross section is approximately a factor~6 smaller than the $AA$
cross section, whereas the $AH$ cross section is
negligible~\cite{lhc_hh}.

\section{A heavier Higgs boson at linear colliders}
\label{sec:lc}

We now turn our attention to Higgs boson pair production in $e^+e^-$
collisions. A detailed study of how well the Higgs boson self-coupling
for $m_H=120$~GeV can be measured in $e^+e^-\to ZHH$ at
$\sqrt{s}=500$~GeV was presented in Ref.~\cite{LC_HH4}. Here we
consider $HH$ production via both $e^+e^-\to ZHH$ and $e^+e^-\to
HH\nu\bar\nu$ for a heavier Higgs boson, in particular for $m_H\geq
150$~GeV, where $H\to WW$ decays dominate; and for center of mass
energies in the range $0.5-1$~TeV. For $HH\nu\bar\nu$ production, we
take into account the $WW$ fusion diagrams considered in
Ref.~\cite{LC_HH1}, as well as the diagrams contributing to $e^+e^-\to
Z(\to\nu\bar\nu)HH$.  The $WW$ fusion diagrams contribute only to the
$HH\nu_e\bar\nu_e$ final state.\medskip

If $m_H\leq 140$~GeV, the dominant decay mode of the SM Higgs boson is
$H\to b\bar{b}$. In this mass range, Higgs bosons which are pair
produced in $e^+e^-\to ZHH$ can be identified with high efficiency via
the $b$-quark content of the system recoiling against the $Z$ boson
(which may either decay hadronically or leptonically). As demonstrated
in Ref.~\cite{LC_HH4}, it is sufficient to require that only one
$b$-quark is tagged. If one assumes a tagging efficiency for
$b$-quarks of $\epsilon_b=0.8$~\cite{LC}, this can be done with an
efficiency close to $100\%$. In $HH\nu\bar\nu$ production, on the
other hand, the presence of neutrinos makes it necessary to fully
reconstruct the event~\cite{LC_HH1}. In this case we require that both
Higgs bosons decay into $b$-quark pairs, and that all $b$-quarks are
identified.

\begin{figure}[t!] 
\begin{center}
\includegraphics[width=13cm]{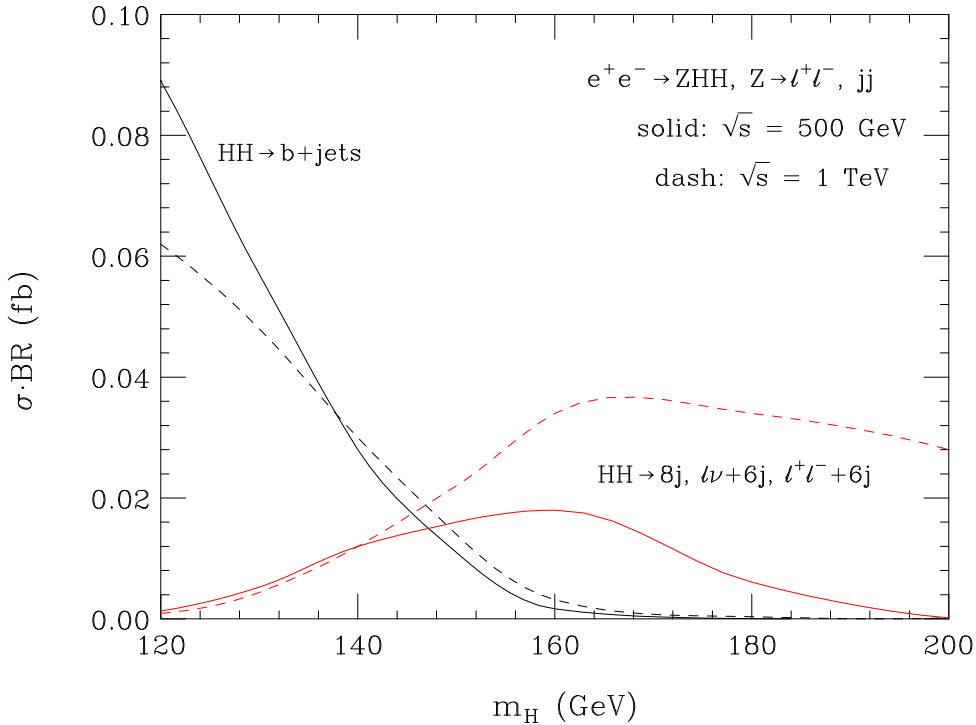}
\vspace*{2mm}
\caption[]{\label{fig:zhh1} The total $e^+e^-\to ZHH$ cross section times
  branching ratio for $\sqrt{s}=500$~GeV (solid lines) and
  $\sqrt{s}=1$~TeV (dashed lines) for various final states. }
\end{center}
\end{figure}

\begin{figure}[t!] 
\begin{center}
\includegraphics[width=13cm]{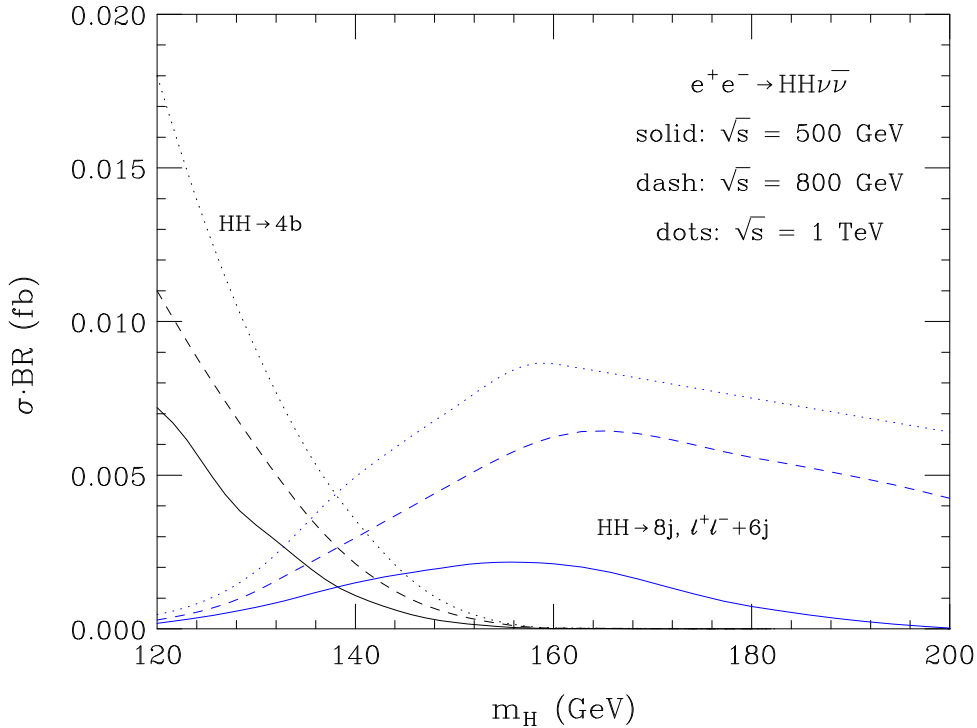}
\vspace*{2mm}
\caption[]{\label{fig:nunuhh1} The $e^+e^-\to HH\nu\bar\nu$ cross 
  section times branching ratio for $\sqrt{s}=500$~GeV (solid lines),
  $\sqrt{s}=800$~GeV (dashed lines) and $\sqrt{s}=1$~TeV (dotted
  lines) for various final states. The curves for $HH\to 4b$ also
  contain the efficiency for tagging four $b$-quarks.}
\end{center}
\end{figure}

For $ZHH$ production, where both Higgs bosons decay either into
$W$ or $Z$ boson pairs, we consider $HH\to 8$~jets, $HH\to
\ell\nu+6$~jets ($\ell=e,\,\mu$) and $HH\to \ell^+\ell^-+6$~jets. The
first two final states have the largest individual branching ratios of
all $4V$ ($V=W,\,Z$) channels. For $e^+e^-\to HH\nu\bar\nu$, we
restrict ourselves to the neutrino-less 8~jets and
$\ell^+\ell^-+6$~jets final states.
The total $e^+e^-\to ZHH$ and $e^+e^-\to HH\nu\bar\nu$ cross sections times
branching ratios for the final states discussed above are shown in
Figs.~\ref{fig:zhh1} and~\ref{fig:nunuhh1} as a function of $m_H$.
Since the $H\to b\bar{b}$ branching ratio decreases rapidly for
increasing $m_H$, the $e^+e^-\to ZHH$, $HH\to b+$jets cross section
falls quickly. For $m_H<140$~GeV, the $ZHH$ cross section at a 1~TeV
linear collider is smaller than that obtained at a machine operating
at 500~GeV. For larger Higgs boson masses, phase space severely limits
the cross section for $\sqrt{s}=500$~GeV. The combined $ZHH$, $HH\to
8$~jets, $\ell\nu+6$~jets, and $\ell^+\ell^-+6$~jets cross section
peaks for $m_H\approx 165$~GeV, with only 18 (35) events/ab$^{-1}$
produced at $\sqrt{s}=500$~GeV ($\sqrt{s}=1$~TeV) before any detection
efficiencies are taken into account. The $e^+e^-\to HH\nu\bar\nu$
cross sections are about a factor~4 to~10 smaller than those for $ZHH$
production for the center of mass energies considered here.  In
addition to $ZHH$ and $HH\nu\bar\nu$ production, there is the process
$e^+e^-\to HHe^+e^-$. Its cross section is a factor $3-5$ smaller than
that for $e^+e^-\to HH\nu\bar\nu$ for values of collider energies and
Higgs boson masses considered here. Thus, we ignore $HHe^+e^-$
production.

The results shown in Figs.~\ref{fig:zhh1} and~\ref{fig:nunuhh1} are
for unpolarized beams. Assuming a polarization of ${\cal P}_-=0.8$ for
the electron and ${\cal P}_+=0.6$ for the positron beam, the
$e^+e^-\to ZHH$ ($e^+e^-\to HH\nu_e\bar\nu_e$) cross section is a
factor~1.70 (2.88) larger than that obtained for unpolarized
beams.\medskip

Since we are interested in determining the Higgs boson self coupling,
we note that the sensitivity to $\lambda$ of both the $ZHH$ and
$HH\nu\bar\nu$ cross sections decreases (increases) with increasing
collider energy (Higgs boson mass).  While the $ZHH$ cross section
grows with rising $\lambda$ in the vicinity of the SM value, the $WW$
fusion cross section diminishes~\cite{LC_HH1}. These effects partially
cancel in the $e^+e^-\to HH\nu\bar\nu$ cross section and considerably
reduce its sensitivity to the Higgs boson self-coupling.

\subsection{$\mathbf m_H\leq 140$~GeV}

Figures~\ref{fig:zhh1} and~\ref{fig:nunuhh1} demonstrate that $ZHH$
production followed by $HH\to b+$jets is the dominant source of $HH$
events in the SM if $m_H\leq 140$~GeV. The main backgrounds in this
channel are top quark and $W$ pair production. These are efficiently
suppressed by performing a neural net (NN) analysis. Such an analysis,
including a detailed detector simulation, was presented in
Ref.~\cite{LC_HH4} for $m_H=120$~GeV and $\sqrt{s}=500$~GeV. It
concluded that $\lambda$ can be determined with a precision of about
$23\%$ if an integrated luminosity of 1~ab$^{-1}$ can be achieved. As
we have seen in Sec.~\ref{sec:lhc}, the limits achievable at hadron
colliders for $m_H=120$~GeV are significantly weaker. To see whether
this statement also holds for other Higgs boson masses (with $m_H\leq
140$~GeV) and other collider energies, it is necessary to extend the
result of Ref.~\cite{LC_HH4} to larger Higgs boson masses and collider
energies.

Since we do not have the tools available which enabled the authors of
Ref.~\cite{LC_HH4} to carry out their analysis, we use the following
simple procedure to estimate bounds for $\lambda_{HHH}$. We calculate
sensitivity limits from the number of $ZHH$, $Z\to\ell\ell,\, jj$,
$HH\to b+$jets ($\ell=e,\,\mu$) signal events and the number of
background events for a NN output parameter of $N_N>0.9$. $N_N$
measures how ``signal-like'' events are with $N_N=1$ ($N_N=0$)
corresponding to perfect signal-like (background-like) events. For
$m_H=120$~GeV, $\sqrt{s}=500$~GeV and an integrated luminosity of
1~ab$^{-1}$, these numbers are taken from Ref.~\cite{LC_HH4}. We
calculate the number of signal events for larger Higgs boson masses
and higher center of mass energies from the $ZHH$ cross section,
assuming that the NN efficiency is independent of both in the ranges
considered.  We estimate the number of background events assuming that
it scales with the top quark cross section as a function of the
collider energy.

A slight complication arises from the functional form of the $ZHH$
cross section, which is a quadratic function of $\lambda_{HHH}$. It is
possible that two separate ranges of $\lambda_{HHH}$ exist which are
consistent with the measured cross section. In this case we select the
range which includes the SM value, $\lambda_{HHH}=1$.\medskip

\begin{figure}[t!] 
\begin{center}
\includegraphics[width=13.cm]{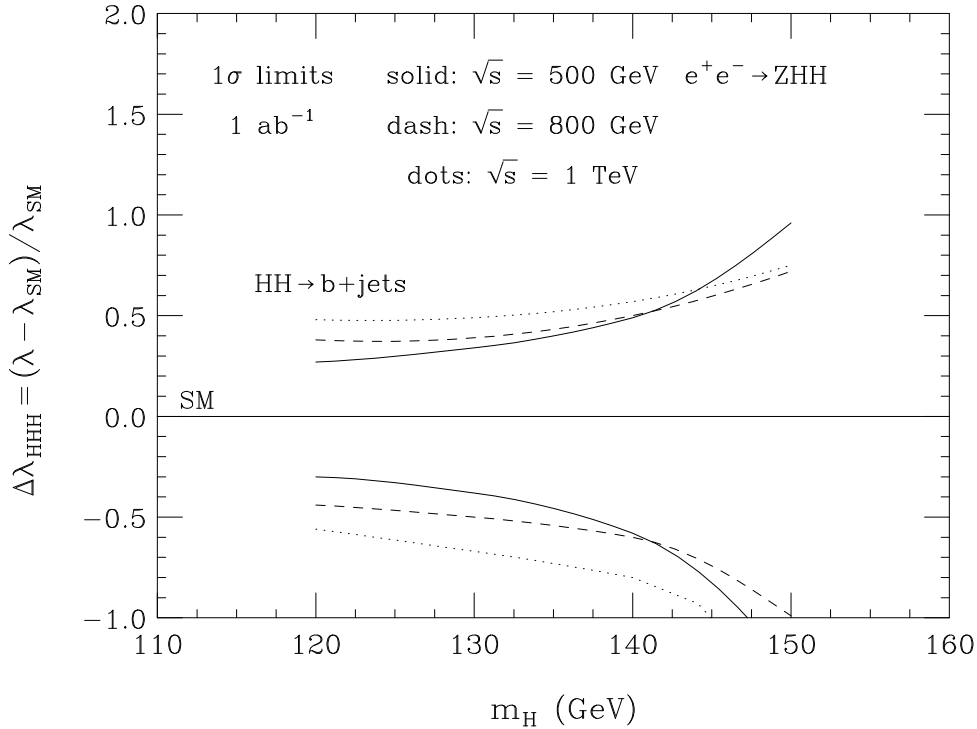}
\vspace*{2mm}
\caption[]{\label{fig:zhh_lim} Estimated $1\sigma$ limits achievable 
  for $\Delta\lambda_{HHH}=(\lambda-\lambda_{SM})/\lambda_{SM}$ in
  $e^+e^-\to ZHH$, $Z\to\ell\ell,\, jj$, $HH\to b+$~jets
  ($\ell=e,\,\mu$) for $\sqrt{s}=500$~GeV (solid lines),
  $\sqrt{s}=800$~GeV (dashed lines), and $\sqrt{s}=1$~TeV (dotted
  lines) and an integrated luminosity of 1~ab$^{-1}$. The allowed
  region is between the two lines of equal texture.}
\end{center}
\end{figure}

We show our results as a function of $m_H$ in Fig.~\ref{fig:zhh_lim}.
For $m_H=120$~GeV and $\sqrt{s}=500$~GeV, we find bounds which are
about a factor~1.2 weaker than those reported in Ref.~\cite{LC_HH4}.
For $\sqrt{s}=800$~GeV and the same $m_H$ our estimated limits agree
well with those found in Ref.~\cite{LC_HH3}. Since the $ZHH$ cross
section and its sensitivity to $\lambda$ decrease with increasing
collider energy, a linear collider operating at 500~GeV offers the
best chance for a precise measurement of $\lambda$ for $m_H\leq
140$~GeV; the bounds we obtain for $\sqrt{s}=500$~GeV are up to a
factor~1.4 (1.9) more stringent than those achievable for
$\sqrt{s}=800$~GeV ($\sqrt{s}=1$~TeV). The advantage of operating at
500~GeV gradually disappears with increasing Higgs boson mass, due to
the reduced phase space. For $\sqrt{s}<500$~GeV, the bounds on
$\lambda$ degrade quickly, likewise due to the rapidly shrinking phase
space.  The sensitivity limits achievable weaken by a factor~1.8 (1.2)
for $\sqrt{s}=500$~GeV ($\sqrt{s}=1$~TeV) if $m_H$ increases from
120~GeV to 140~GeV; for $m_H=140$~GeV one will not be able to probe
$\lambda$ with a precision of better than $50\%$ for unpolarized
beams. Since the bounds which could be obtained from $pp\to
b\bar{b}\tau^+\tau^-$ degrade by a similar amount in this range (see
Sec.~\ref{sec:lhc}), we conclude that a $0.5-1$~TeV linear collider
offers a significantly better chance to probe $\lambda$ for the mass
range from 120~GeV to 140~GeV. If both the electron and positron beams
can be polarized, the bounds derived here improve by a factor~1.3,
assuming $80\%$ polarization for the electron beam and $60\%$ for the
positron beam, and the same integrated luminosity as for unpolarized
beams.

\subsection{$\mathbf m_H>140$~GeV}
\label{sec:lc_heavy}

If $m_H>140$~GeV, the channels yielding the largest event rates are
$e^+e^-\to ZHH$ with $Z\to jj$ and $HH\to 8$~jets or $\ell\nu+6$~jets.
Channels where one of the Higgs bosons decays into $b\bar{b}$, as well
as $HH\nu\bar\nu$ production, result in negligible cross sections. In
this section, we therefore concentrate on the $\ell\nu+8$~jet and the
10~jet final states.

Final states of similar structure and complexity are encountered in
$t\bar{t}H$ production. If $H\to WW$, one also expects $\ell\nu+8$~jet
and 10~jet events~\cite{gay}. If the Higgs boson predominantly decays
to bottom quarks, $\ell\nu+6$~jet and 8~jet events are produced. In
contrast to $ZHH$ production, the $\ell\nu+$jets and all jets events
originating from $t\bar{t}H$ production contain two or more
$b$-quarks. The processes $e^+e^-\to t\bar{t}H\to
q\bar{q}b\,\ell\nu\bar{b}\,b\bar{b}, q\bar{q}b\,q\bar{q}b\,b\bar{b}$
were analyzed in detail in Ref.~\cite{juste}.

The main background processes contributing both to $e^+e^-\to ZHH$ and
$e^+e^-\to t\bar{t}H$ are $WW+$~jets, $t\bar{t}+$jets and QCD multijet
production. In the $t\bar{t}H$ case, the combined background cross
section is several orders of magnitude larger than that of the signal.
To reduce the background, one first imposes preselection cuts to
remove as much background as possible. One optimizes $S/B$ via a NN
analysis in a second step. We list the efficiencies and signal to
background ratios found in Ref.~\cite{juste} for $t\bar{t}H$
production for both steps in the analysis in Table~\ref{tab:two}.
\begin{table}
\caption{Efficiencies, $\epsilon$, and signal to background ratios,
$S/B$, obtained in Ref.~\protect\cite{juste} for $e^+e^-\to t\bar{t}H$.}   
\label{tab:two}
\vskip 3.mm
\begin{tabular}{c|cc|cc}
 & \multicolumn{2}{c}{Cuts analysis} & \multicolumn{2}{c}{NN
analysis} \\
final state & $\epsilon$ & $S/B$ & $\epsilon$ & $S/B$\\
\tableline
$t\bar{t}H\to\ell\nu+6$~jets & 0.54 & 0.03 & 0.27 & 0.5\\
$t\bar{t}H\to 8$~jets & 0.77 & 0.03 & 0.085 & 0.9
\end{tabular}
\end{table}

Before imposing any cuts, the $e^+e^-\to ZHH\to 10$~jets
($\ell\nu+8$~jets) cross section is about a factor~30 smaller than
that for $e^+e^-\to t\bar{t}H\to 8$~jets ($\ell\nu+6$~jets). Due to
the additional two jets in the final state, the background to $ZHH$
production is suppressed by a factor $\alpha_s^2$, resulting in a
background cross section which is roughly one order of magnitude
smaller than in the $t\bar{t}H$ case. The signal to background ratios
before cuts for $ZHH$ and $t\bar{t}H$ production therefore are
similar, and a NN analysis for $ZHH$ production will likely lead to
reductions of the signal efficiencies and the background rates which
are similar to those encountered in the $t\bar{t}H$ analysis of
Ref.~\cite{juste}.

Exact sensitivity bounds for $\lambda$ in $e^+e^-\to
ZHH\to\ell\nu+8$~jets and $e^+e^-\to ZHH\to 10$~jets could be derived
only after performing a detailed NN analysis, which is beyond the
scope of this paper. Instead, we investigate how the sensitivity
bounds for $\lambda$ depend on the signal efficiencies and the signal
to background ratio.  We then explore the prospects for determining
the Higgs boson self-coupling in $ZHH$ production for $m_H>140$~GeV
using the results of Table~\ref{tab:two}, which we argued may be used
as rough guidelines.\medskip

\begin{figure}[t!] 
\begin{center}
\includegraphics[width=13.cm]{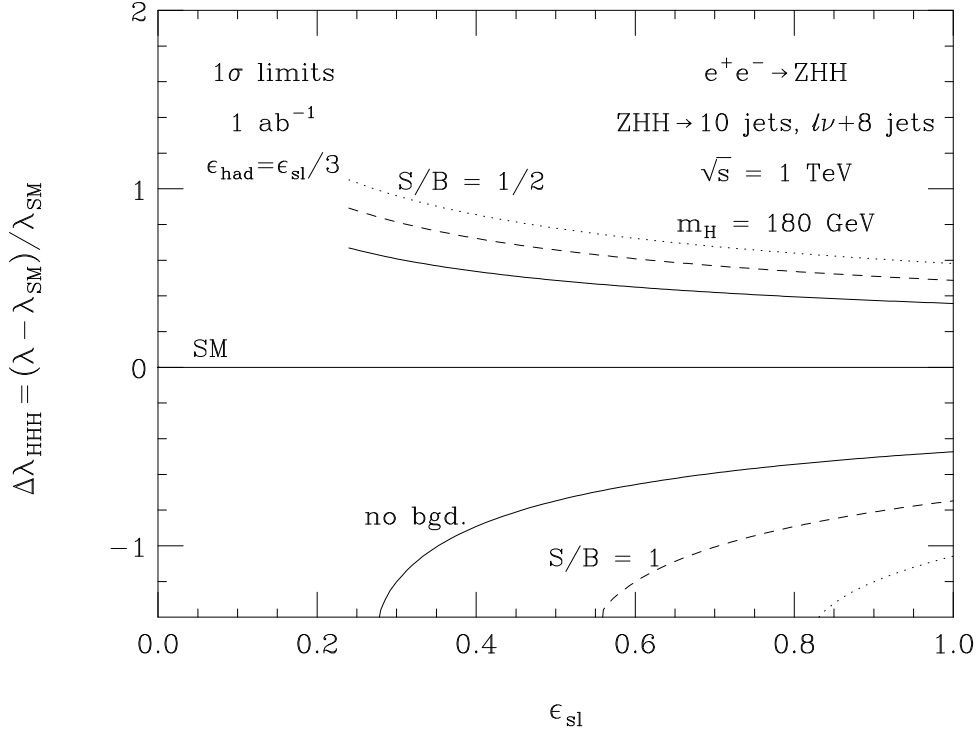}
\vspace*{2mm}
\caption[]{\label{fig:lc_eff1} Estimated $1\sigma$ limits achievable 
  for $\Delta\lambda_{HHH}=(\lambda-\lambda_{SM})/\lambda_{SM}$ in
  $e^+e^-\to ZHH\to \\ 10$~jets, $\ell\nu+8$~jets for $\sqrt{s}=1$~TeV
  and an integrated luminosity of 1~ab$^{-1}$, as a function of the
  detection efficiency of the semi-leptonic $\ell\nu+8$~jet final
  state, $\epsilon_{sl}$. The ratio of the efficiencies of the
  hadronic 10~jet and the semi-leptonic $\ell\nu+8$~jet channels is
  assumed to be $\epsilon_{had}/\epsilon_{sl}=1/3$. The solid curves
  represent the limits if no background is present. The dashed and
  dotted lines display the $1\sigma$ limits if $S/B=1$ and $S/B=1/2$,
  respectively. The allowed region is between the two lines of equal
  texture.}
\end{center}
\end{figure}

We perform our analysis assuming $m_H=180$~GeV, $\sqrt{s}=1$~TeV, and
an integrated luminosity of 1~ab$^{-1}$. Since the number of signal
events is small, we combine the $\ell\nu+8$~jet and 10~jet final
states and use the total cross section (including branching ratios and
efficiencies) to derive sensitivity limits.  We show the $1\sigma$
sensitivity limits for $\Delta\lambda_{HHH}$ in Fig.~\ref{fig:lc_eff1}
as a function of the efficiency of the semi-leptonic $\ell\nu+8$~jet
final state, $\epsilon_{sl}$, for several choices of $S/B$, assuming a
fixed ratio of $\epsilon_{had}/\epsilon_{sl}=1/3$ for the efficiencies
of the hadronic 10~jet and the semi-leptonic $\ell\nu+8$~jet channels.
This $\epsilon_{had}/\epsilon_{sl}$ is motivated by the NN results of
Ref.~\cite{juste} (see Table~\ref{tab:two}).  For larger (smaller)
values of $\epsilon_{had}/\epsilon_{sl}$, somewhat more (less)
stringent bounds are obtained.

\begin{figure}[t!]
\begin{center}
\includegraphics[width=13.cm]{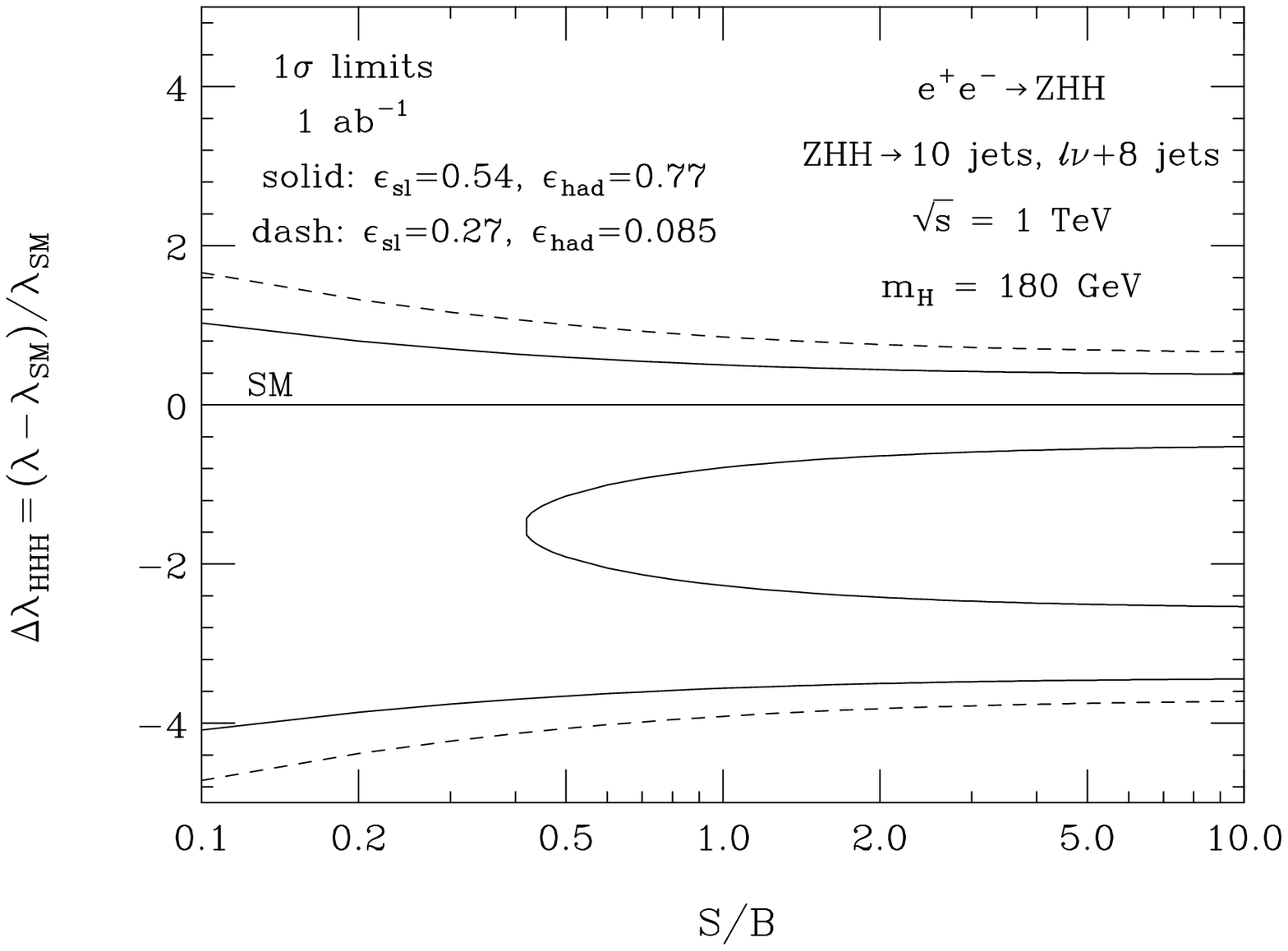}
\vspace*{2mm}
\caption[]{\label{fig:lc_sb} Estimated $1\sigma$ limits achievable 
  for $\Delta\lambda_{HHH}=(\lambda-\lambda_{SM})/\lambda_{SM}$ in
  $e^+e^-\to ZHH\to \\ 10~{\rm jets}$, $\ell\nu+8$~jets for
  $\sqrt{s}=1$~TeV and an integrated luminosity of 1~ab$^{-1}$, as a
  function of the signal to background ratio $S/B$. The solid curves
  represent the limits for $\epsilon_{sl}=0.54$ and
  $\epsilon_{had}=0.77$. The dashed curves display the bounds for
  $\epsilon_{sl}=0.27$ and $\epsilon_{had}=0.085$. The allowed region
  is between the two lines of equal texture.}
\end{center}
\end{figure}

Figure~\ref{fig:lc_eff1} demonstrates that the bounds achievable on
$\lambda$ in $e^+e^-\to ZHH\to 10$~jets and $e^+e^-\to
ZHH\to\ell\nu+8$~jets depend strongly on $\epsilon_{sl}$ and $S/B$.
The latter dependence is more transparent in Fig.~\ref{fig:lc_sb},
where we show the $1\sigma$ sensitivity limits as a function of $S/B$
for the $t\bar{t}H$ preselection and NN efficiencies (see
Table~\ref{tab:two}).
Sensitivity bounds better than unity occur only for high efficiencies,
similar to the $t\bar{t}H$ preselection efficiencies, and if
$S/B>0.5$. For the more likely case that $\epsilon_{sl}$,
$\epsilon_{had}$ and $S/B$ are similar to the values obtained in the
$t\bar{t}H$ analysis, a first-generation LC could obtain only very
loose bounds on $\lambda_{HHH}$. Using the values for the NN analysis
listed in Table~\ref{tab:two} for illustration purposes, one finds
\begin{equation}
\label{eq:lambda_lc}
-4.1 < \Delta\lambda_{HHH} < 1.0
\end{equation}
at the $1\sigma$ level for $m_H=180$~GeV, $\sqrt{s}=1$~TeV and
1~ab$^{-1}$. We find very similar constraints for both options of
Table~\ref{tab:two}, the cuts analysis and the NN analysis. To be
sure, one should really perform a NN analysis for $HH$ that anti-tags
$b$-jets, instead of tags them to confirm the presence of top quarks.
However, as we argued previously, $S/B$ is already quite poor, and
other backgrounds are of the same size as $t\bar{t}$, so the values in
Table~\ref{tab:two} can be taken to be fair approximations to what one
might expect for the Higgs pair production signal. As noted earlier in
this section, the limits achievable improve by about a factor~1.3 for
electron and positron beam polarizations of $80\%$ and $60\%$,
respectively, and if the same integrated luminosity as in the
unpolarized case can be reached.\medskip

For comparison, the LHC (SLHC) $\Delta\lambda_{HHH}$ can give
$1\sigma$ constraints of $-0.3<\Delta\lambda_{HHH}<1.6$
($-0.10<\Delta\lambda_{HHH}<0.12$)~\cite{BPR} for $m_H=180$~GeV.  The
LHC with 300~fb$^{-1}$ will thus be able to better constrain
$\Delta\lambda_{HHH}$ for negative values than a first generation
linear collider with 1~ab$^{-1}$ operating at 1~TeV. This is a
fortuitous effect of the destructive interference of the two diagrams
in the gluon fusion process.  For positive values, the linear collider
may enjoy a slight advantage over the LHC. We reach similar
conclusions for $m_H=160$~GeV and $\sqrt{s}=800$~GeV. For
$m_H<160$~GeV and $m_H>180$~GeV, fewer than 5 signal events would be
seen if efficiencies are smaller than 0.5, disallowing bounds to be
placed on $\lambda_{HHH}$.

\section{Reconstructing the Higgs Potential at Lepton and Hadron
  Colliders}
\label{sec:pot}

The results of the previous sections, together with those of
Refs.~\cite{LC_HH3,LC_HH4} and~\cite{BPR}, can be used to compare the
capabilities of future lepton and hadron colliders to reconstruct the
Higgs potential. In order to translate bounds on
$\Delta\lambda_{HHH}=(\lambda-\lambda_{SM})/\lambda_{SM}$ into
constraints on the Higgs potential which can be graphically displayed,
it is convenient to consider the scaled Higgs potential
\begin{equation}
\label{eq:scale_pot}
{2\over v^2m_H^2}\,V(x)=x^2+\lambda_{HHH}x^3+
{1\over 4}\,\tilde\lambda_{4H}x^4~ \, ,
\end{equation}
where
\begin{equation}
x={\eta_H\over v}~,
\end{equation}
$\tilde\lambda_{4H}=\tilde\lambda/\lambda_{SM}$ is the four Higgs
boson self-coupling normalized to the SM value ($\lambda_{SM}$ is
given in Eq.~(\ref{eq:lamsm})), $\eta_H$ is the physical Higgs field,
and $v=(\sqrt{2}G_F)^{-1/2}$ is the vacuum expectation value. In the
following we assume $\tilde\lambda_{4H}=1$.\medskip

\begin{figure}[t!]
\begin{center}
\includegraphics[width=13.cm]{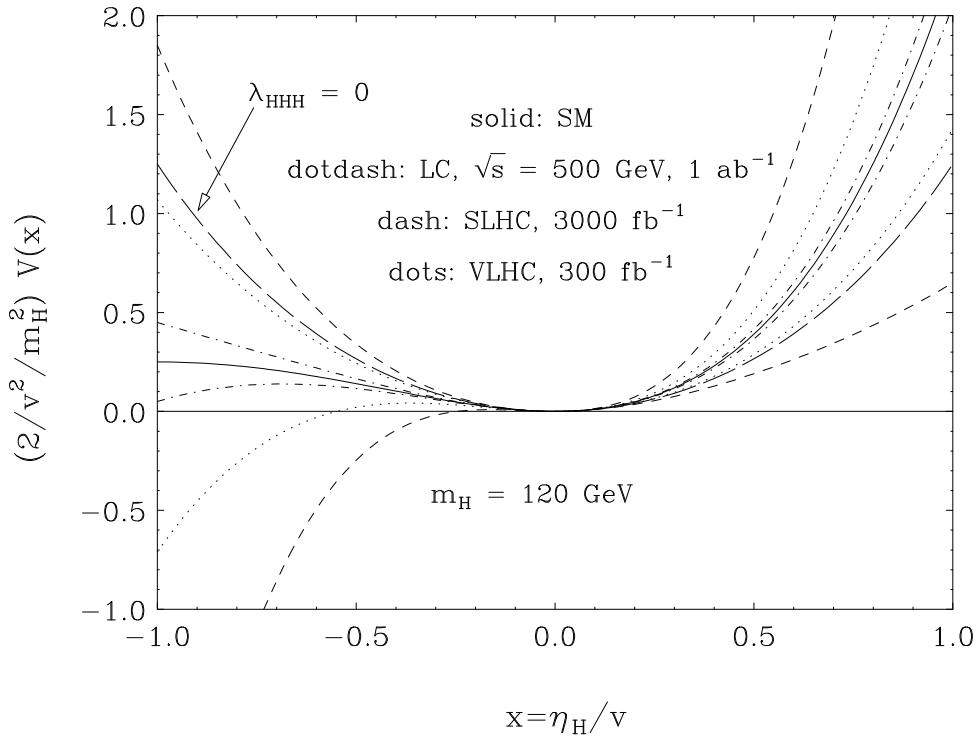}
\vspace*{2mm}
\caption[]{\label{fig:pot1} Constraints on the scaled Higgs potential
  for $m_H=120$~GeV. The dashed (dotted) lines show the limits
  achievable at the SLHC (VLHC) in the $b\bar{b}\tau\tau$ channel.
  The dot-dashed curves are derived using the limits of
  Ref.~\cite{LC_HH4} for $e^+e^-\to ZHH$, $Z\to\ell\ell,\, jj$, $HH\to
  b+$jets, $\sqrt{s}=500$~GeV, and an integrated luminosity of
  1~ab$^{-1}$. The allowed region is between the two lines of equal
  texture. The solid line represents the SM Higgs potential, and the
  long-dashed line shows the result for a vanishing Higgs boson
  self-coupling.}
\end{center}
\end{figure}

In Fig.~\ref{fig:pot1} we show how well the scaled Higgs potential can
be reconstructed for a light Higgs boson of mass $m_H=120$~GeV.
As demonstrated in Sec.~\ref{sec:lhc}, hadron colliders have only very
limited capabilities to probe $\lambda$ if $m_H\leq 140$~GeV. Higgs
boson pair production with dual $H\to b\bar{b}$ decays is swamped by
the QCD $4b$ background. A slightly better chance is offered if one of
the Higgs bosons decays into $\tau$ pairs, with one $\tau$ lepton
decaying leptonically and the second decaying into hadrons. This
channel will be invisible at the LHC, due to the small signal cross
section. At the SLHC and VLHC a sufficient number of signal events is
expected. However, the QCD $b\bar{b}\tau\tau$ and $t\bar{t}$
backgrounds limit the sensitivity to $\Delta\lambda_{HHH}$ to ${\cal
  O}(1)$ and the Higgs potential can be reconstructed only poorly
(dashed and dotted lines in Fig.~\ref{fig:pot1}). In contrast, at a
500~GeV linear collider with an integrated luminosity of 1~ab$^{-1}$,
$\Delta\lambda_{HHH}$ can be measured with a precision of about
$20\%$~\cite{LC_HH4}, and the Higgs potential can be reconstructed
fairly accurately.

We draw similar conclusions for other Higgs boson masses in the range
$120~{\rm GeV}<m_H<140$~GeV; the limits achievable for
$\Delta\lambda_{HHH}$ both at lepton and hadron colliders gradually
weaken by about a factor~2 if $m_H$ is increased from 120~GeV to
140~GeV.  While the constraints on the Higgs potential improve with
increasing machine energy for hadron colliders, the opposite is true
in $e^+e^-$ collisions. Here, both the $ZHH$ cross section and its
sensitivity to the Higgs boson self-coupling decrease with increasing
values of the collider energy.

\begin{figure}[t!]
\begin{center}
\includegraphics[width=13.cm]{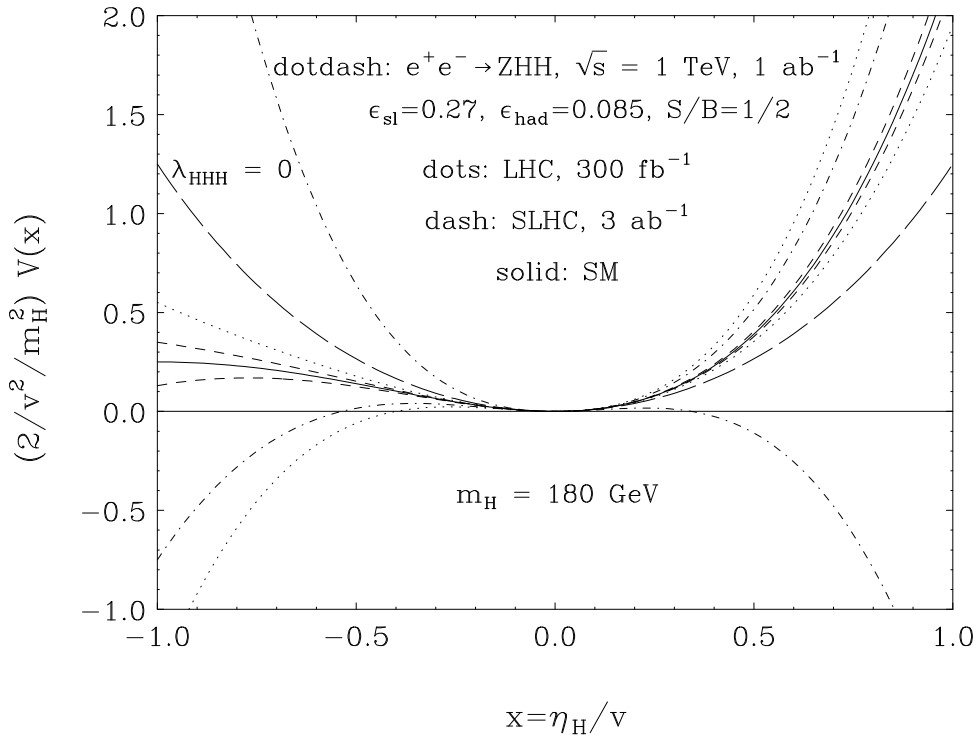}
\vspace*{2mm}
\caption[]{\label{fig:pot2} Constraints on the scaled Higgs potential
  for $m_H=180$~GeV. The dashed (dotted) lines show the limits which
  can be achieved at the SLHC (LHC) in the
  $(jj\ell^\pm\nu)\,(jj{\ell'}^\pm\nu)$ channel~\cite{BPR}.  The
  dot-dashed curves are derived using the limits of
  Eq.~(\ref{eq:lambda_lc}) for $e^+e^-\to ZHH\to 10$~jets,
  $\ell\nu+8$~jets, $\sqrt{s}=1$~TeV, and an integrated luminosity of
  1~ab$^{-1}$. These limits {\sl assume} an efficiency of
  $\epsilon_{sl}=0.27$ ($\epsilon_{had}=0.085$) for
  $ZHH\to\ell\nu+8$~jets ($ZHH\to 10$~jets), and a signal to
  background ratio of $S/B=1/2$.  The allowed region is between the
  two lines of equal texture. The solid line represents the SM Higgs
  potential, and the long-dashed line shows the result for a vanishing
  Higgs boson self-coupling.}
\end{center}
\end{figure}

If the Higgs boson decays predominantly into a pair of $W$-bosons,
i.e. if $m_H\geq 150$~GeV, a completely different picture emerges.
Figure~\ref{fig:pot2} displays how well the Higgs potential may be
reconstructed at future colliders if $m_H=180$~GeV.
The dashed and dotted lines show the constraints on the Higgs
potential which one may hope to achieve at the SLHC (LHC) in the
$(jj\ell^\pm\nu)\,(jj{\ell'}^\pm\nu)$ channel.  We derived these
curves by converting the $95\%$ CL limits of Ref.~\cite{BPR} for
$\Delta\lambda_{HHH}$ into $1\sigma$ limits and using
Eq.~(\ref{eq:scale_pot}). While LHC experiments will only be able to
put mild constraints on $V(x)$, a luminosity upgrade of the LHC will
make it possible to reconstruct the Higgs potential quite precisely
for this $m_H$ range.\medskip

At a linear collider with a center of mass energy in the $0.8-1$~TeV
range and an integrated luminosity of 1~ab$^{-1}$, the number of Higgs
boson pair events is very limited. The dominant $WW+$~jets and
$t\bar{t}\:+$~jets backgrounds are several orders of magnitude larger
than the signal. As discussed in Sec.~\ref{sec:lc_heavy}, any analysis
which attempts to improve $S/B$ to an acceptable level is likely to
significantly reduce the signal efficiencies, and thus the sensitivity
to $\Delta\lambda_{HHH}$. As a result, it will be difficult to
constrain the Higgs potential using linear collider data if $m_H\geq
150$~GeV. This point is illustrated by the dot-dashed lines in
Fig.~\ref{fig:pot2}, which show how poorly $V(x)$ is constrained via
$e^+e^-\to ZHH\to 10$~jets, $\ell\nu+8$~jets, at a 1~TeV $e^+e^-$
collider with an integrated luminosity of 1~ab$^{-1}$, if the
efficiencies and the signal to background ratios would be equal to
those obtained in Ref.~\cite{juste} for $e^+e^-\to t\bar{t}H$. Similar
results are obtained for a wide range of efficiencies and $S/B$ values
(see Fig.~\ref{fig:lc_sb}); the dot-dashed lines in
Fig.~\ref{fig:pot2} thus are representative.  We obtain results
similar to those shown in Fig.~\ref{fig:pot2} for $m_H=160$~GeV. For
Higgs boson masses between 150~GeV and 160~GeV, and for $m_H>180$~GeV,
there are not enough signal events at an $e^+e^-$ collider with center
of mass energy in the $0.8-1$~TeV range to constrain the Higgs
potential.

\begin{figure}[t!]
\begin{center}
\includegraphics[width=13.cm]{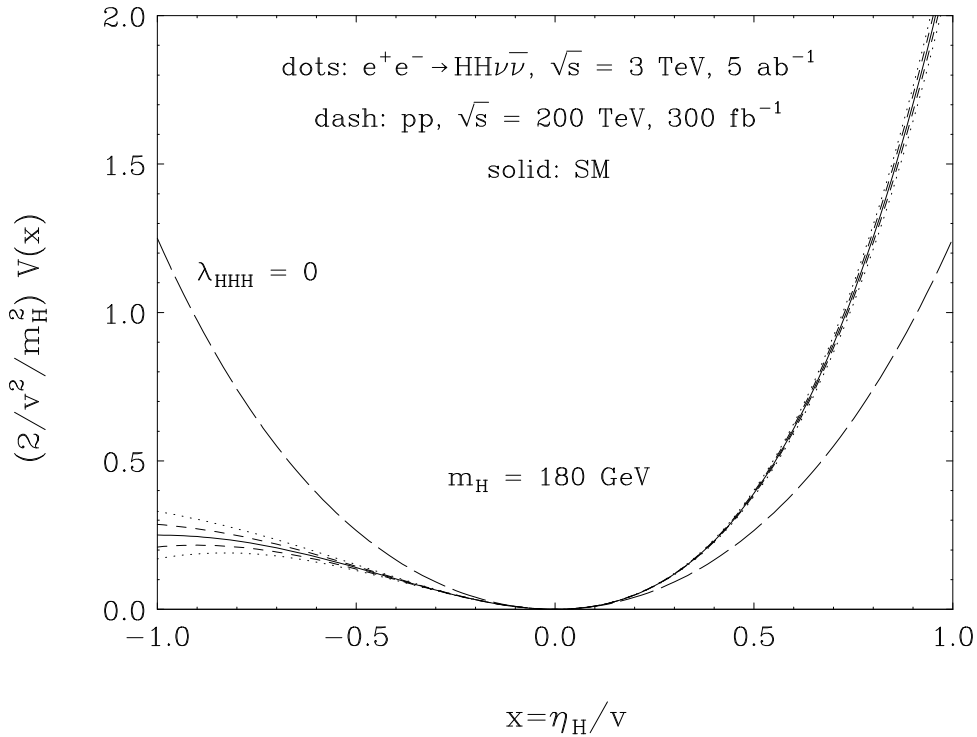}
\vspace*{2mm}
\caption[]{\label{fig:pot3} Constraints on the scaled Higgs potential
  for $m_H=180$~GeV. The dashed lines show the limits achievable at
  the VLHC in the $(jj\ell^\pm\nu)(jj{\ell'}^\pm\nu)$ channel with
  300~fb$^{-1}$~\cite{BPR}. The dotted use the limits of
  Ref.~\cite{LC_HH3} for $e^+e^-\to HH\nu\bar\nu$, $HH\to 4W$ at CLIC
  ($\sqrt{s}=3$~TeV, $\int\!{\cal L}dt=5~{\rm ab}^{-1}$).  The allowed
  region is between the two lines of equal texture. The solid line
  represents the SM Higgs potential, and the long-dashed line shows
  the result for a vanishing Higgs boson self-coupling.}
\end{center}
\end{figure}

It should be noted that the prospects to determine the Higgs boson
self-coupling and to reconstruct the Higgs potential at an $e^+e^-$
collider for a Higgs boson with mass larger than 150~GeV improve
dramatically at larger energies. The $e^+e^-\to HH\nu\bar\nu$ cross
section grows rapidly with energy~\cite{LC_HH1}, reaching about 0.5~fb
for $m_H=180$~GeV and $\sqrt{s}=3$~TeV, the energy of the two beam
linear collider CLIC proposed by CERN~\cite{CLIC}. At such a machine,
with an integrated luminosity of 5~ab$^{-1}$, it should be possible to
determine $\lambda_{HHH}$ with a precision of about
$8\%$~\cite{LC_HH3}. The VLHC could achieve similar or better
precision~\cite{BPR}. As mentioned in Sec.~\ref{sec:sec2}, one-loop
electroweak radiative corrections change $\lambda_{HHH}$ by a similar
amount~\cite{yuan}. At CLIC or a VLHC it will thus be possible to
probe the Higgs boson self-coupling at the quantum level. The
constraints on the shape of the Higgs potential from $HH\nu\bar\nu$
production at CLIC and $gg\to HH\to 4W\to
(jj\ell^\pm\nu)(jj{\ell'}^\pm\nu)$ at a VLHC are shown in
Fig.~\ref{fig:pot3}.
It will be possible to accurately reconstruct the Higgs potential at
these machines.

\section{Discussion and Conclusions}
\label{sec:conc}

A direct experimental investigation of the Higgs potential represents
a conclusive test of the mechanism of electroweak symmetry breaking
and mass generation. After the discovery of an elementary Higgs boson
and the test of its couplings to fermions and gauge bosons,
experimental evidence that the shape of the Higgs potential has the
form required for breaking the electroweak symmetry will complete the
proof that the masses of fermions and weak bosons are generated by
spontaneous symmetry breaking. To probe the shape of the Higgs
potential, one must determine the Higgs boson self-coupling.\medskip

Only Higgs boson pair production at lepton or hadron colliders can
measure the Higgs boson self-coupling. Ref.~\cite{LC_HH4} carried out
a detailed study of how well this could be done for $m_H=120$~GeV in
$ZHH$ production at a 500~GeV $e^+e^-$ collider.  Refs.~\cite{LC_HH3}
and~\cite{LC_HH2} considered Higgs boson pair production at $e^+e^-$
colliders operating in the $2-5$~TeV range.
Refs.~\cite{SLHC,BPR,blondel} determined the prospects at hadron
colliders for $150~{\rm GeV}<m_H<200$~GeV. In this paper, we tried to
fill in gaps in the existing literature by considering Higgs boson
pair production for a light Higgs boson with mass $m_H\leq 140$~GeV at
hadron colliders, and for a Higgs boson of mass $m_H>120$~GeV at
$e^+e^-$ colliders, with particular emphasis on the range $m_H\geq
150$~GeV where decays into $W$ pairs dominate.

For pair production of a light Higgs boson at hadron colliders we
considered the dominant $4b$ final state and the $b\bar{b}\tau\tau$
channel. The $4b$ final is swamped by the QCD background, which is
more than two orders of magnitude larger than the signal. At the LHC,
the number of $b\bar{b}\tau\tau$ signal events is too small to yield
any useful information on the Higgs boson self-coupling. At the SLHC
and VLHC, however, the improved signal to background ratio does yield
somewhat better limits on $\lambda$ than the $4b$ final state,
although the signal cross section in the $b\bar{b}\tau\tau$ channel is
significantly smaller. Performing a $\chi^2$ analysis for the visible
invariant mass distribution, we found that it will be difficult to
probe the Higgs boson self-coupling to better than about one, even at
a 200~TeV VLHC.

If $m_H\leq 140$~GeV, the Higgs boson self-coupling can be determined
with much greater precision in $e^+e^-\to ZHH$, $HH\to b+$jets. We
extrapolated the results of Ref.~\cite{LC_HH4} to $m_H>120$~GeV and
higher center of mass energies and found that, since both the $ZHH$
cross section and its sensitivity to $\lambda$ decrease with
increasing center of mass energy, a 500~GeV $e^+e^-$ collider
operating is optimally suited to probe the Higgs boson self-coupling
for $120~{\rm GeV}\leq m_H\leq 140$~GeV. The limits on the Higgs boson
self-coupling for $e^+e^-$ collisions at $\sqrt{s}=500$~GeV, assuming
an integrated luminosity of 1~ab$^{-1}$, are typically a factor 5 (10)
more stringent than those that would come from a VLHC (SLHC) in this
mass range. Data from a 500~GeV linear collider, however, will not be
sufficiently sensitive to probe the electroweak one-loop corrections
to $\lambda$. A multi-TeV $e^+e^-$ collider will be the only machine
capable of this if $m_H\leq 140$~GeV~\cite{LC_HH3}.

Due to phase space restrictions, a center of mass energy of at least
800~GeV would be needed to search for Higgs pair production in
$e^+e^-$ collisions if $m_H\geq 150$~GeV. For $\sqrt{s}=0.8-1$~TeV,
$e^+e^-\to ZHH\to 10~{\rm jets},\,\ell\nu+8$~jets via Higgs boson
decays into weak boson pairs are the dominant Higgs pair production
channels. The main contributions to the background originate from
$t\bar{t}\:+$~jets and $WW+$~jets production, with cross sections
several orders of magnitude larger than the signal. In such a
situation, the only hope to improve the signal to background ratio,
$S/B$, to an acceptable level is a NN analysis. We studied how the
sensitivity bounds on $\lambda$ depend on the signal efficiencies and
$S/B$. Using the results for $e^+e^-\to t\bar{t}H$
production~\cite{juste}, where final states of similar complexity and
$S/B$ are encountered, as guidelines, we conclude that it will be
difficult to determine the Higgs boson self-coupling at a linear
collider with $\sqrt{s}=0.8-1$~TeV with a precision equal to that
which can be reached at the LHC with 300~fb$^{-1}$. We reach this
conclusion for a broad range of efficiencies and $S/B$ values;
therefore, it does {\sl not} depend on the specific values which were
used. Experiments at both a multi-TeV $e^+e^-$ collider (where
$HH\nu\bar\nu$ production is the main source of Higgs pair events) and
a VLHC will be able to probe the one-loop electroweak radiative
corrections to $\lambda$ for $m_H\geq 150$~GeV.\medskip

Our results show that hadron colliders and $e^+e^-$ linear
colliders with $\sqrt{s}\leq 1$~TeV are complementary: for $m_H\leq
140$~GeV, linear colliders offer far better prospects in measuring the
Higgs boson self-coupling, $\lambda$; for a Higgs boson in the range
$m_H\geq 150$~GeV, the opposite is true. However, to actually perform
a meaningful measurement at a hadron collider would demand precision
Higgs boson properties input from a linear collider for the top quark
Yukawa coupling, the $HWW$ coupling, and the total Higgs boson decay
width.\medskip

Finally, we have explored how well various future colliders may
constrain the shape of the Higgs potential, $V(\eta_H)$. To visualize
how a non-standard Higgs self-coupling affects $V(\eta_H)$, we
introduced a scaled version of the potential, expressed in terms of
the dimensionless ratio $x=\eta_H/v$ (see Eq.~(\ref{eq:scale_pot})).
Results for several machines and choices of $m_H$ are shown in
Figs.~\ref{fig:pot1}~--~\ref{fig:pot3}.

\acknowledgements

We would like to thank K.~Desch, P.~Gay, K.~Jakobs, A.~Juste, A.~Nikitenko,
T.~Stelzer, A.~Turcot, G.~Wilson and P.M.~Zerwas for useful
discussions.  One of us (U.B.) would like to thank the Fermilab Theory
Group, where part of this work was carried out, for its generous
hospitality.  This research was supported in part by the National
Science Foundation under grants No.~PHY-9970703 and No.~PHY-0139953.



\bibliographystyle{plain}

\end{document}